\documentclass[aps,prb,showpacs,groupedaddress,amssymb,twocolumn]{revtex4}
\usepackage{bm,amsmath}
\usepackage[dvips]{graphicx,color}

\begin{document}

\title{Mesoscopic electron and phonon transport through a curved wire}

\author{Shi-Xian Qu and Michael R. Geller}

\affiliation{Department of Physics and Astronomy, University of Georgia, Athens, Georgia 30602-2451}

\date{April 9, 2004}

\begin{abstract}
There is great interest in the development of novel nanomachines that use charge, spin, or energy transport, to enable new sensors with unprecedented measurement capabilities. Electrical and thermal transport in these mesoscopic systems typically involves wave propagation through a nanoscale geometry such as a quantum wire. In this paper we present a general theoretical technique to describe wave propagation through a curved wire of uniform cross-section and lying in a plane, but of otherwise arbitrary shape. The method consists of $(i)$ introducing a local orthogonal coordinate system, the arclength and two locally perpendicular coordinate axes, dictated by the shape of the wire; $(ii)$ rewriting the wave equation of interest in this system; $(iii)$ identifying an effective scattering potential caused by the local curvature; and $(iv)$, solving the associated Lippmann-Schwinger equation for the scattering matrix. We carry out this procedure in detail for the scalar Helmholtz equation with both hard-wall and stress-free boundary conditions, appropriate for the mesoscopic transport of electrons and (scalar) phonons. A novel aspect of the phonon case is that the reflection probability always vanishes in the long-wavelength limit, allowing a simple perturbative (Born approximation) treatment at low energies. Our results show that, in contrast to charge transport, curvature only barely suppresses thermal transport, even for sharply bent wires, at least within the two-dimensional scalar phonon model considered. Applications to experiments are also discussed.
\end{abstract}


\maketitle
\clearpage

\section{INTRODUCTION}

A new class of nanomachines is attempting to measure extremely minute amounts of energy, of the order a few neV, and to use such calorimeters to probe fundamental properties of thermal conduction in the nanoscale regime. \cite{Tighe etal,Roukes phonon counting,Schwab etal quantization,Yung etal quantization,Cleland book}  Like the related case of electrical conduction,\cite{Beenakker review,Datta book} low-temperature thermal conduction in nanostructures is entirely different than in macroscopic materials because the phonons are in the mesoscopic regime, where they scatter elastically but not inelastically. Because inelastic scattering is required to establish thermodynamic equilibrium, there is a breakdown of Fourier's law and the heat equation, which assume a local thermodynamic equilibrium characterized by a spatially varying temperature profile. These nanodevices have inspired considerable theoretical work on thermal transport by phonons in the mesoscopic limit.\cite{Nishiguchi etal,Rego and Kirczenow,Angelescu etal,Blencowe landauer,Kambili etal,Leitner etal molecule,Leitner and Wolynes nanocrystal,Blencowe and Vitelli information,Ozpineci and Ciraci,Cross and Lifshitz,Glavin,Santamore and Cross,Patton and Geller weak link,Cleland etal phononic,Sun etal landauer}

\begin{figure}
\includegraphics[width=10.0cm]{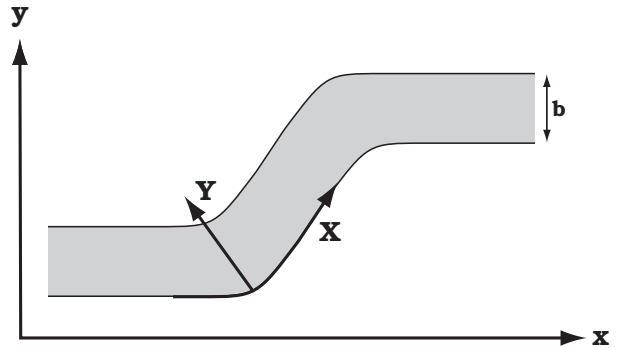}
\caption{\label{bentwire figure} An example of the type of two-dimensional curved wire geometry considered in this paper. The shape of the wire is used to define local orthogonal coordinates $X$ and $Y$, with $X$ the arclength along one of the edges. The width of the wire is $b$.}
\end{figure}

In this paper we introduce a general method to calculate the scattering matrix for waves propagating through a curved wire or waveguide. The wire is assumed to be of uniform cross-section and lying in a plane, but the curved segment may have any smooth curvature profile,\cite{smoothness footnote} such as that shown schematically in Fig.~\ref{bentwire figure}. The ends of the wire (the ``leads") are also assumed to be straight. For definiteness we consider two-dimensional waves described by the scalar Helmholtz equation
\begin{equation}
\big[ \nabla^2 + \alpha \big] \Phi({\bf r}) = 0, \ \ \ \ \ {\bf r} \equiv (x,y)
\label{original wave equation}
\end{equation}
which is appropriate for electrons or scalar phonons in flat wires with rectangular cross-section.\cite{thickness footnote} Here $\alpha(\epsilon) \equiv 2 m \epsilon/\hbar^2$ in the case of electrons of energy $\epsilon$ and mass $m$, whereas $\alpha(\epsilon) \equiv \epsilon^2/\hbar^2 v^2$ in the case of scalar phonons of energy $\epsilon$ and bulk sound velocity $v$.  Electron spin is of no importance here and is neglected. The boundary conditions at the edges of the wire are
\begin{eqnarray}
\Phi({\bf r}) &=& 0, \ \ \ \ \ ({\rm for \ electrons}) \cr
{\bf n} \cdot {\bm \nabla} \Phi({\bf r}) &=& 0, \ \ \ \ \ ({\rm for \ phonons}) 
\label{original boundary conditions}
\end{eqnarray}
where ${\bf n}({\bf r})$ is a local outward-pointing normal. Here we have assumed conventional hard-wall boundary conditions for the electronic states, but stress-free conditions for the elastic waves because in this case the wires are usually freely suspended.

Although the wave equation in Eq.~(\ref{original wave equation}) is certainly simple, the scattering problem described here is complicated because the boundary conditions (\ref{original boundary conditions}) are applied along the curves defining the two edges of the wire. Our approach involves rewriting Eqs.~(\ref{original wave equation}) and (\ref{original boundary conditions}) in terms of new curvilinear coordinates $X$ and $Y$, dictated by the shape of the wire, such that the wave equation becomes more complicated (the wire's curvature produces an effective potential), but the boundary conditions become trivial. We choose $X$ to be the arclength along one edge of the wire, and $Y$ is locally perpendicular. This transformation allows us to apply the standard techniques of scattering theory, including solution of the Lippmann-Schwinger equation, in the $X Y$ frame. 

A particularly novel aspect of the phonon transport problem is that the reflection probability always vanishes in the long-wavelength limit, permitting an analytic (second-order Born approximation) treatment at low energies. The energy-dependent transmission probability is then expressed as as a simple functional of the curvature profile $\kappa(X)$, making possible a straightforward analysis of a variety of wire shapes. The fact that long-wavelength phonons have perfect transmission is a consequence of the rigid-body nature of the underlying system: An elastic wave with infinite wavelength is just an adiabatic rigid translation of the wire, which must transmit energy perfectly.

There has been considerable attention given to mesoscopic electron transport through curved wires and waveguides,\cite{Londergan book} but none to thermal transport. Electron transmission probabilities in curved wires are usually obtained by mode-matching, a method restricted to piecewise separable geometries (wires composed of straight segments, circles, and other shapes where the wave equation is separable). A related problem that has been studied extensively is the formation of electronic bound states and resonances in curved wires, where the mapping to local curvilinear coordinates is also often used.\cite{Londergan book,Exner and Seba,Goldstone and Jaffe} Surprisingly, we are not aware of any work using moving frames and then directly solving the resulting Lippmann-Schwinger equation in that basis.\cite{constant width footnote} Nor are we aware of the use of this method in the extensive microwave engineering literature,\cite{Lewin book,Collin book} where the (more generally applicable but purely numerical) finite-element method is the technique of choice.

In the next section we carry out the above analysis for the two-dimensional Helmholtz equation. In Sec.~\ref{electron section} we consider electron transport through a circular right-angle bend, recovering results obtained by Sols and Macucci \cite{Sols and Macucci} and by Lin and Jaffe\cite{Lin and Jaffe} using mode-matching methods. Our main results are given in Sec.~\ref{phonon section}, where we address thermal transport through curved wires. Sec.~\ref{discussion section} contains a discussion of our conclusions and the experimental implications of this work.

\section{APPLICATION TO SCALAR WAVE EQUATION} 

We now explain our method in detail and apply it to the scalar scattering problem stated in Eqs.~(\ref{original wave equation}) and (\ref{original boundary conditions}).

\subsection{Curvilinear coordinate system}

First we use the shape of the wire to define a curvilinear coordinate system, the arclength $X$ along one edge and a locally perpendicular coordinate $Y$. Which edge one chooses is of course arbitrary. The direction of increasing $Y$ will be chosen so that $X, Y,$ and $z$ form a right-handed coordinate system.  Both edges are assumed to be a smooth plane curves.\cite{smoothness footnote}

The unit tangent vector 
\begin{equation}
{\bf e}_X \equiv {d{\bf r} \over dX} = {dx \over dX} \, {\bf e}_x +  {dy \over dX} \, {\bf e}_y 
\label{eX definition}
\end{equation}
and normal
\begin{equation}
{\bf e}_Y \equiv -{dy \over dX} \, {\bf e}_x +  {dx \over dX} \, {\bf e}_y
\label{eY definition}
\end{equation}
define local orthonormal basis vectors for the $XY$ frame. The sign in Eq.~(\ref{eY definition}) is chosen so that ${\bf e}_X \times {\bf e}_Y = {\bf e}_z$. We then use the Frenet-Serret equation
\begin{equation}
{d {\bf e}_X  \over d X} = \kappa(X) \, {\bf e}_Y
\label{Frenet-Serret equation}
\end{equation} 
to define a signed curvature $\kappa(X)$ of the $Y \! = \! 0$ edge. According to this definition, $\kappa(X)$ is positive when $Y$ increases toward the center of curvature. 
We will also make use of the metric tensor
\begin{equation}
g = \left[ \begin{matrix} (1-\kappa Y)^2 & 0 \cr 0 & 1 \end{matrix} \right]
\label{metric}
\end{equation}  
in the $X Y$ system. 

\subsection{Helmholtz equation in curvilinear coordinates}

Next we make a coordinate transformation from ${\bf r}$ to ${\bf R} \! \equiv \! (X,Y)$, and rewrite the wave equation in terms of these coordinates. A convenient way to do this is to use the identity 
\begin{equation}
\nabla^2 = ({\rm det} \, g)^{-{1 \over 2}} {\partial \over \partial X_i} ({\rm det} \, g)^{{1 \over 2}} \, g_{ij}^{-1} \,  {\partial \over \partial X_j}.
\label{Beltrami identity}
\end{equation}
The Helmholtz equation (\ref{original wave equation}) then becomes
\begin{equation}
\big[ \partial_X^2 + \partial_Y^2 - V + \alpha \big] \Phi({\bf R}) = 0,
\label{new wave equation}
\end{equation}
where we have separated the combination $\partial_X^2 + \partial_Y^2$ from the many terms that appear on the right-hand-side of Eq.~(\ref{Beltrami identity}), and combined the remaining ones into an effective potential $V$. The potential $V$ is itself a differential operator; an explicit expression will be given below. The boundary conditions (\ref{original boundary conditions}) are now
\begin{eqnarray}
\Phi({\bf R}) &=& 0, \ \ \ \ \ ({\rm for \ electrons}) \cr
\partial_Y  \Phi({\bf R}) &=& 0, \ \ \ \ \ ({\rm for \ phonons}) 
\label{new boundary conditions}
\end{eqnarray}
on the edges $Y \! = \! 0$ and $Y \! = \! b$, with $b$ the width of the wire. A scattering state $\Phi({\bf R})$ becomes fully determined once we specify its behavior as $X \rightarrow \pm \infty$.

\begin{figure}
\includegraphics[width=8.5cm]{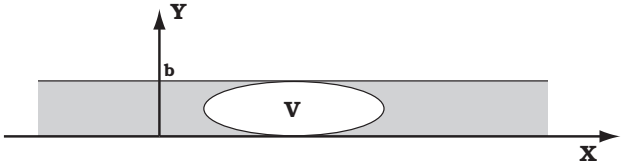}
\caption{\label{straightwire figure} Scattering problem in the $X Y$ frame. Here the wire appears straight, but the curvature induces an effective potential $V$ that causes scattering.}
\end{figure}

In the $X Y$ frame the wire appears straight, as illustrated in Fig.~\ref{straightwire figure}. The scattering potential vanishes in the leads because the wire is straight there. We are now able to use conventional scattering theory.

\subsection{Unperturbed scattering states}

\begin{figure}
\includegraphics[width=7.5cm]{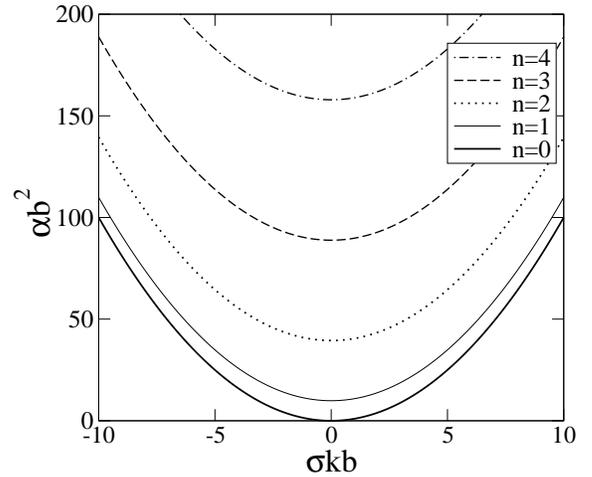}
\caption{\label{dispersion figure} Dispersion relation for unperturbed scattering states, for both spinless electrons and scalar phonons. $\alpha$ is equal to $2 m \epsilon/\hbar^2$ in the case of electrons, or $\epsilon^2/\hbar^2 v^2$ in the case of phonons. $\sigma$ is $1$ on the right half of the figure, and $-1$ on the left. $b$ is the wire width.}
\end{figure}

The unperturbed ($V \! = \! 0$) scattering states for both spinless electrons and scalar phonons are labeled by three quantum numbers $\sigma$, $n$, and $\epsilon$, and can be written in the form
\begin{equation}
\phi_{\sigma n \epsilon}({\bf R}) = c_{n \epsilon} \, e^{\sigma i k_{n \epsilon} X} \chi_n({\bf Y}),
\label{unperturbed states}
\end{equation}
where
\begin{equation}
k_{n \epsilon} \equiv \sqrt{\alpha(\epsilon) - (n \pi /b)^2}
\label{dispersion relation}
\end{equation}
is the wave number along the wire,
\begin{equation}
\chi_n(Y) \equiv \begin{cases} \sqrt{2/b} \, \sin(n \pi Y/b) & \text{for electrons} \\ 
\sqrt{(2 \! - \! \delta_{n0})/b} \, \cos(n \pi Y/b) &  \text{for phonons} \end{cases}
\label{transverse wave function}
\end{equation}
is a trigonometric function satisfying the transverse boundary conditions of Eq.~(\ref{new boundary conditions}), and
\begin{equation}
c_{n \epsilon} \equiv {1 \over \sqrt{2 \pi}} \bigg| {\partial k_{n \epsilon} \over \partial \epsilon} \bigg|^{1/2}
\end{equation}
is a real normalization constant. $\sigma$ is a chirality index, defined by
\begin{equation}
\sigma = \begin{cases} +1 & \text{if moving in $+X$ direction} \\ -1 &  \text{if moving in $-X$ direction} \end{cases}
\end{equation}
and $n$ is an integer-valued branch index. The transverse eigenfunctions $\chi_n(Y)$ are normalized according to
\begin{equation}
\int_0^b dY \ \chi_n(Y) \,  \chi_{n'}(Y) = \delta_{nn'} .
\end{equation}
The dispersion relation given by Eq.~(\ref{dispersion relation}) is shown in Fig.~\ref{dispersion figure}.

The allowed values of the quantum numbers $\sigma$, $n$, and $\epsilon$ are as follows: The allowed energies form a continuum, from $\epsilon_{\rm min}$ to $\infty$. Here
\begin{equation}
\epsilon_{\rm min} \equiv \begin{cases}  {\textstyle{\hbar^2 \pi^2 \over 2 m b^2}}  & \text{for electrons} \\  0 &  \text{for phonons} \end{cases}.
\label{minimum energy}
\end{equation}
For each value of energy, the branch index takes the values in a set $S$ of integers defined by
\begin{equation}
S \equiv \begin{cases}  1, 2, \dots, N & \text{for electrons} \\  0, 1, 2, \dots, N &  \text{for phonons} \end{cases}
\label{allowed n values}
\end{equation}
where 
\begin{equation}
N(\epsilon) \equiv \sum_{n=0}^\infty \Theta\big[\alpha(\epsilon) - {\textstyle{n^2 \pi^2 \over b^2}} \big] \ - \ 1,
\label{N definition}
\end{equation}
with $\Theta(x)$ the unit step function. For electrons, $N$ is the number of propagating channels below energy $\epsilon$, whereas for phonons the number of propagating channels is $N+1$. Finally, for each allowed value of $\epsilon$ and $n$, $\sigma$ takes on the values $\pm 1$.

The free scattering states satisfy the orthonormality and completeness conditions
\begin{equation}
\int d^2R \ \phi_{\sigma n \epsilon}^*({\bf R}) \,  \phi_{\sigma' \! n' \! \epsilon'}({\bf R}) =  \delta_{\sigma \sigma'} \, \delta_{nn'} \, \delta(\epsilon - \epsilon')
\end{equation}
and
\begin{equation}
\int_{\epsilon_{\rm min}}^\infty \! \! d\epsilon \! \sum_{\sigma = \pm 1} \sum_{n \in S} \phi_{\sigma n \epsilon}^*({\bf R}) \, \phi_{\sigma n \epsilon}({\bf R}') = \delta({\bf R}-{\bf R}'),
\end{equation}
where $n$ takes the values given in Eq.~(\ref{allowed n values}).

\subsection{Effective potential}

The terms $\partial_X^2 + \partial_Y^2$ have been separated out from the right-hand-side of Eq.~(\ref{Beltrami identity}) so that the eigenfunctions $\Phi({\bf R})$ reduce to that of a straight wire when $V=0$. Accordingly, the effective potential is given by
\begin{equation}
V = {\kappa^2 Y^2 - 2 \kappa Y  \over (1-\kappa Y)^2} \, \partial^2_X - {\kappa' Y \over (1-\kappa Y)^3} \, \partial_X + {\kappa \over 1-\kappa Y} \, \partial_Y.
\label{effective potential}
\end{equation}
There will be no singularities in $V$ as long as
\begin{equation} 
-\infty < \kappa b  < 1.
\label{curvature condition}
\end{equation}
The condition (\ref{curvature condition}) guarantees that both the $Y \! = \! 0$ and $Y \! = \! b$ edges of the wire are smooth.

In applications where the radius of curvature $|\kappa|^{-1}$ is much larger than $b$, Eq.~(\ref{effective potential}) can be simplified. To leading order in $\kappa b$ the effective potential reduces to
\begin{equation}
V = - 2 \kappa Y \, \partial^2_X - \kappa' Y (1+ 3 \kappa Y) \, \partial_X + \kappa \, \partial_Y.
\label{small curvature potential}
\end{equation}

\subsection{Lippmann-Schwinger equation}

The scattering problem in the $XY$ frame can be solved by standard methods. The Lippmann-Schwinger equation for an eigenfunction $\Phi({\bf R})$ with (electron or phonon) energy $\epsilon$ is 
\begin{equation}
\Phi({\bf R}) = \phi_{\rm in}({\bf R}) + \int \! d^2R' \, G_0({\bf R},{\bf R}',\epsilon) \, V \, \Phi({\bf R}'),
\label{lippmann-schwinger equation}
\end{equation}
where $\phi_{\rm in}({\bf R})$ is a free scattering state coming in from the left, and where $G_0({\bf R},{\bf R}', \epsilon)$ is the Green's function for the unperturbed Helmholtz equation, satisfying
\begin{equation}
\big[ \partial_X^2 + \partial_Y^2 + \alpha(\epsilon) \big] G_0({\bf R}, {\bf R}', \epsilon) = \delta({\bf R} - {\bf R}'),
\end{equation}
along with the boundary condition that $G_0({\bf R}, {\bf R}', \epsilon)$ vanishes as $ |X - X'| \rightarrow \infty.$ The Lippmann-Schwinger equation gives the solution of Eq.~(\ref{new wave equation}) subject to the condition that $\Phi({\bf R})$ reduces to the incident state $\phi_{\rm in}({\bf R})$ when $X \rightarrow -\infty$. 

The unperturbed Green's function for both electrons and phonons is
\begin{equation}
G_0({\bf R}, {\bf R}',\epsilon) = - {i \over 2} \sum_{n=0}^\infty  {\chi_n(Y) \, \chi_n(Y') \over k_{n \epsilon} } \ e^{i k_{n \epsilon} |X-X'|}.
\label{G0} 
\end{equation}
We note that the summation in Eq.~(\ref{G0}) is {\it not} restricted to the values given in Eq.~(\ref{allowed n values}). In particular, off-shell values of $k_{n \epsilon}$ are included. Furthermore, in the electron case the $n=0$ term in the summation vanishes, because the transverse eigenfunction $\chi_0(Y)$ vanishes.

We will also need to write Eq.~(\ref{lippmann-schwinger equation}) in the alternative form
\begin{equation}
\Phi({\bf R}) = \phi_{\rm in}({\bf R}) + \int \! d^2R' \ G({\bf R},{\bf R}',\epsilon) \, V \phi_{\rm in}({\bf R}'), 
\label{alternative lippmann-schwinger equation}
\end{equation}
where $G({\bf R},{\bf R}',\epsilon )$ is the Green's function of the {\it perturbed} Helmholtz equation, satisfying
\begin{eqnarray}
G({\bf R},{\bf R}',\epsilon)&=& G_0({\bf R},{\bf R}',\epsilon)  \nonumber \\
&+& \int d^2R^{''} G_0({\bf R},{\bf R}'',\epsilon) \, V \, G({\bf R}'',{\bf R}',\epsilon). \ \ \ \ \ \ \ 
\label{Dyson equation}
\end{eqnarray}

\subsection{Transmission probability}

The transmission coefficient matrix $t_{nn'}(\epsilon)$ gives the probability amplitude for a right-moving electron or phonon to forward scatter from branch $n$ to branch $n'$ at energy $\epsilon$. We define $t_{nn'}(\epsilon)$ to be zero if one or both branches have minima above $\epsilon$.

We can obtain a formal expression for $t_{nn'}(\epsilon)$ by writing the $X \rightarrow \infty$ limit of the unperturbed Green's function as
\begin{equation}
G_0({\bf R}, {\bf R}',\epsilon) \rightarrow  - {i \over 2} \sum_{n \in S} {\phi_{{\rm R} n \epsilon}({\bf R}) \, 
\phi_{{\rm R} n \epsilon}^*({\bf R}^\prime) \over k_{n \epsilon} c_{n\epsilon}^2} ,
\label{asymptotic G0} 
\end{equation}
where the subscripts R denote right-moving ($\sigma = +1$) waves. The summation in Eq.~(\ref{asymptotic G0}) is now restricted to $n \le N$ because the higher lying contributions are exponentially small in the $X \rightarrow \infty$ limit. Then from Eqs.~(\ref{lippmann-schwinger equation}) and (\ref{asymptotic G0}) we obtain
\begin{equation}
\lim_{X \rightarrow \infty} \! \Phi({\bf R}) = \phi_{{\rm R}n_{\rm i}\epsilon}({\bf R}) - {i \over 2} \sum_{n \in S} 
{\langle \phi_{{\rm R}n\epsilon}|V|\Phi \rangle \over k_{n \epsilon} c_{n\epsilon}^2} \, \phi_{{\rm R}n\epsilon}({\bf R})
\end{equation}
where the incoming right-moving wave is assumed to be in channel $n_{\rm i}$. Therefore we conclude that
\begin{equation}
\lim_{X \rightarrow -\infty} \! \Phi = \phi_{{\rm R}n_{\rm i}\epsilon}
\end{equation}
and
\begin{equation}
\lim_{X \rightarrow \infty} \! \Phi = \sum_{n \in S}  t_{n_{\rm i} n}(\epsilon) \, \phi_{{\rm R}n\epsilon},
\end{equation}
where
\begin{equation}
t_{nn'}(\epsilon) \equiv \delta_{nn'} - {i \over 2} {\langle \phi_{{\rm R}n'\epsilon}|V|\Phi \rangle \over k_{n' \epsilon} c_{n'\epsilon}^2} 
\label{transmission matrix}
\end{equation}
is the amplitude for an incident wave $({\rm R}, n ,\epsilon)$ to forward-scatter to $({\rm R}, n' ,\epsilon)$. $t_{nn'}(\epsilon)$ is called
the transmission matrix. 

We emphasize that for the case of electron transport, $t(\epsilon)$ is an $N \! \times \! N$ matrix, where $N$ varies with energy as indicated in Eq.~(\ref{N definition}). For phonons, $t(\epsilon)$ is $N \! + \! 1$-dimensional.

The expectation value in Eq.~(\ref{transmission matrix}) involves the exact scattering state $\Phi$ with boundary condition corresponding to an incident state $({\rm R}, n , \omega)$. Using Eq.~(\ref{alternative lippmann-schwinger equation}) we write this as
\begin{eqnarray}
&&\langle \phi_{{\rm R}n'\epsilon}|V|\Phi \rangle = \langle \phi_{{\rm R}n'\epsilon}|V|\phi_{{\rm R}n\epsilon} \rangle \nonumber \\
&+& \int \phi^*_{{\rm R}n'\epsilon}({\bf R}) \, V_{\! \scriptscriptstyle R} \, G({\bf R},{\bf R}^\prime,\epsilon) \, V_{\! \scriptscriptstyle {R^\prime}} \, 
\phi_{{\rm R}n\epsilon}({\bf R}^\prime), \ \ \ \ \ 
\label{T matrix}
\end{eqnarray}
where $V_{\! \scriptscriptstyle R}$ and $V_{\! \scriptscriptstyle {R^\prime}}$ act on ${\bf R}$ and ${\bf R}^\prime$, respectively. 
The numerical method we use consists of expressing Eq.~(\ref{Dyson equation}) in a basis of unperturbed scattering states, solving
this equation by matrix inversion, and using Eq.~(\ref{T matrix}) to obtain the transmission matrix in Eq.~(\ref{transmission matrix}).

\begin{figure}
\includegraphics[width=4.0cm]{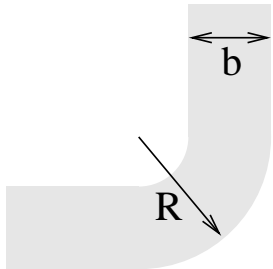}
\caption{\label{angle figure}Section of quantum wire with circular right-angle bend.}
\end{figure}

Finally, we define the energy-dependent transmission probabilities ${\sf T}_{\! \rm el}$ and ${\sf T}_{\! \rm ph}$ for electrons and phonons that determine the electrical and thermal currents. For electrons, the relevant quantity is the ratio of transmitted to incident charge current, given by
\begin{equation}
{\sf T}_{\! \rm el}(\epsilon) = \sum_{n,n'=1}^N \big|t_{nn'}\big|^2 = {\rm Tr} \, t^\dagger t.
\label{electron probability definition}
\end{equation}
For thermal transport by phonons, the relevant quantity is the fraction of transmitted energy current,\cite{Blencowe landauer} given by
\begin{equation}
{\sf T}_{\! \rm ph}(\epsilon) = \sum_{n,n'=0}^N {{\sf v}_{n'} \over {\sf v}_{n}} \, \big|t_{nn'}\big|^2,
\label{phonon probability definition}
\end{equation}
where ${\sf v}_n(\epsilon)$ is the phonon {\it group} velocity in the straight wire, which can be written as $\hbar v^2 k_{n\epsilon} / \epsilon.$ Because the bulk sound velocity $v$ is a constant here, and the scattering is elastic, we can equivalently write Eq.~(\ref{phonon probability definition}) as  
\begin{equation}
{\sf T}_{\! \rm ph}(\epsilon) = \sum_{n,n'=0}^N {k_{n'\epsilon} \over k_{n\epsilon}} \, \big|t_{nn'}\big|^2,
\label{alternative phonon probability definition}
\end{equation}
with the $k_{n\epsilon}$ given by Eq.~(\ref{dispersion relation}).

\subsection{Landauer formula}

The charge current $I$ and linear conductance 
\begin{equation}
G \equiv \lim_{V \rightarrow 0} {I \over V} 
\end{equation}
are related to the electron transmission probability ${\sf T}_{\! \rm el}(\epsilon)$ through the Landauer formula\cite{Beenakker review,Datta book}
\begin{equation}
I = {e \over 2 \pi \hbar} \int_0^\infty \! \! d \epsilon \,  {\sf T}_{\! \rm el}(\epsilon) \, \big[ n_{\rm F}^{\mu_{\rm l}}(\epsilon) 
- n_{\rm F}^{\mu_{\rm r}}(\epsilon) \big].
\label{charge Landauer formula}
\end{equation}
Here $n_{\rm F}^{\mu_{\rm l}}(\epsilon)$ and $n_{\rm F}^{\mu_{\rm r}}(\epsilon)$ are the Fermi distribution functions in the left (l) and right (r) leads, with chemical potentials $\mu_{\rm l}$ and  $\mu_{\rm r}$ differing in proportion to the applied voltage $V = (\mu_{\rm l} - \mu_{\rm r})/e$.

\begin{figure}
\includegraphics[width=7.5cm]{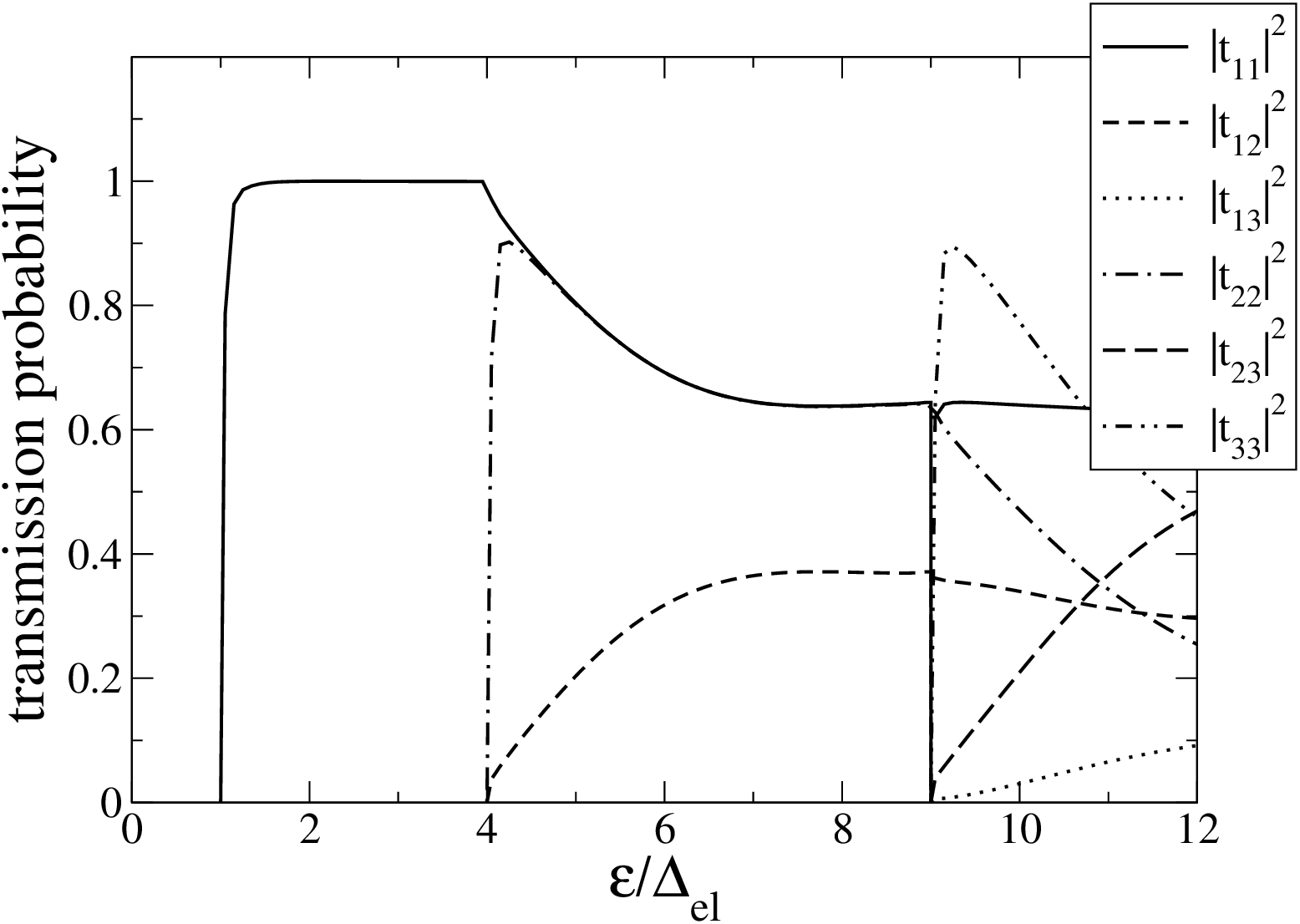}
\caption{\label{Tij-el-R1.2 figure}Individual electron transmission probabilities $|t_{nn'}|^2$ as a function of energy for a circular right-angle bend with $R/b=1.20$. Here $\Delta_{\rm el} \equiv \pi^2 \hbar^2 / 2 m b^2$.}
\end{figure}

Similarly, the thermal current $I_{\rm th}$ and conductance
\begin{equation}
G_{\rm th} \equiv \lim_{\Delta T \rightarrow 0} {I_{\rm th} \over \Delta T} 
\end{equation}
are determined by the phonon transmission probability ${\sf T}_{\! \rm ph}(\epsilon)$ according to\cite{Rego and Kirczenow,Angelescu etal,Blencowe landauer}
\begin{equation}
I_{\rm th} = {1 \over 2 \pi \hbar} \int_0^\infty \! \! d \epsilon \, \epsilon \, {\sf T}_{\! \rm ph}(\epsilon) \, \big[ n_{\rm B}^{T_{\rm l}}(\epsilon) 
- n_{\rm B}^{T_{\rm r}}(\epsilon) \big].
\end{equation}
$n_{\rm B}^{T_{\rm l}}(\epsilon)$ and $n_{\rm B}^{T_{\rm r}}(\epsilon)$ are Bose distribution functions in the left and right leads, with temperatures $T_{\rm l}$ and $T_{\rm r}$.

\section{ELECTRON TRANSPORT}
\label{electron section}

We turn now to an application of our method to coherent electron transport through a circular right-angle bend with outer radius $R$ and width $b$, as shown schematically in Fig.~\ref{angle figure}. In this case the curvature profile is
\begin{equation}
\kappa(X) = {1 \over R} \, \Theta(X) \, \Theta({\textstyle{\pi R \over 2}} - X),
\label{right angle profile}
\end{equation}
where $\Theta$ is the unit step function. The origin of the $X$ coordinate is taken to be one of the locations where the straight and curved sections of the wire meet.

As explained above, the numerical method we use to calculate $t_{nn'}(\epsilon)$ requires the matrix elements $\langle \phi_{\sigma n \epsilon}|V|\phi_{\sigma' n' \epsilon'} \rangle$ of the effective potential (\ref{effective potential}) in the unperturbed scattering states (\ref{unperturbed states}). However, matrix elements of the second term in Eq.~(\ref{effective potential}), which contain the curvature gradient 
\begin{equation}
\kappa^\prime(X) = {1 \over R} \big[ \delta(X)  - \delta(X-{\textstyle{\pi R \over 2}}) \big],
\label{right angle profile derivative}
\end{equation}
involves integrals of a delta function $\delta$ times functions $F(\Theta)$ of $\Theta$. Integrals of this form, involving products of generalized functions, have to be evaluated carefully, as we show in Appendix \ref{delta function appendix}. Apart from this technicality, the application of our method to this geometry is straightforward.

\begin{figure}
\includegraphics[width=7.0cm]{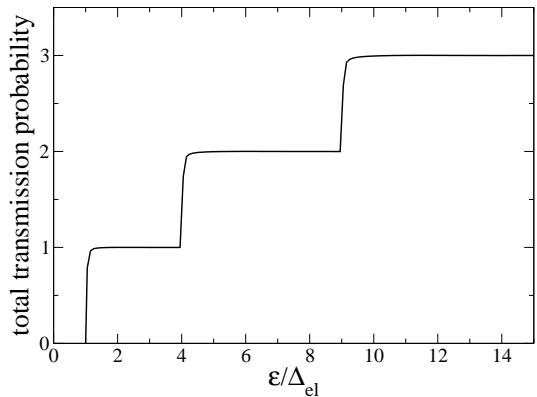}
\caption{\label{T-el-R1.2 figure} Total electron transmission probability ${\sf T}_{\! \rm el}$. System parameters are the same as in Fig.~\ref{Tij-el-R1.2 figure}.}
\end{figure}

Mesoscopic charge transport through bent wires has already been studied extensively, \cite{Londergan book} and we will only consider one case of this, namely $R = 1.2 \, b$. First we calculate the individual electron transmission probabilities $|t_{nn'}(\epsilon)|^2$ from Eq.~(\ref{transmission matrix}) for the lowest few channels $n$ and $n'$. The results are given in Fig.~\ref{Tij-el-R1.2 figure} and are in excellent agreement with the mode-matching results of Sols and Macucci \cite{Sols and Macucci} for the same value of $R/b$ (see Fig.~2a of Ref.~\onlinecite{Sols and Macucci}). Our result for $|t_{11}(\epsilon)|^2$ also agrees qualitatively with that calculated by Lin and Jaffe\cite{Lin and Jaffe} for a right-angle bend with a slightly larger curvature (see Fig.~8 of Ref.~\onlinecite{Lin and Jaffe}).

The total electron transmission probability ${\sf T}_{\! \rm el}(\epsilon)$, defined in Eq.~(\ref{electron probability definition}), is presented in Fig.~\ref{T-el-R1.2 figure} for the same curved wire. At energies given by
\begin{equation}
\epsilon = n^2 \Delta_{\rm el}  \ \ \ \ \ {\rm with} \ \ \ \ \  n=1,2,3,\dots,
\label{electron threshold energies}
\end{equation}
where $\Delta_{\rm el}  \equiv \pi^2 \hbar^2 / 2 m b^2$, additional channels in the wire become propagating and contribute to the transmission probability. The threshold energies (\ref{electron threshold energies}) follows from Eq.~(\ref{dispersion relation}) and Fig.~\ref{dispersion figure}. The principal effect of the curvature in the wire is to soften the transitions at these thresholds. 

\section{PHONON TRANSPORT}
\label{phonon section}

We turn now to the main emphasis of our work, the calculation of transmission probabilities for two-dimensional scalar phonons with energy $\epsilon = \hbar \omega$ and (bulk) sound velocity $v$ to propagate through curved wires. We are not aware of any previous work on this problem.

As before, we consider a circular right-angle bend with outer radius $R$ and width $b$, as shown schematically in Fig.~\ref{angle figure}, with curvature profile given by Eq.~(\ref{right angle profile}). The method of solution is the same as that outlined in Sec.~\ref{electron section} and Appendix \ref{delta function appendix}, except that the transverse parts of the unperturbed scattering states, defined in Eq.~(\ref{transverse wave function}), are now different.

\begin{figure}
\includegraphics[width=7.2cm]{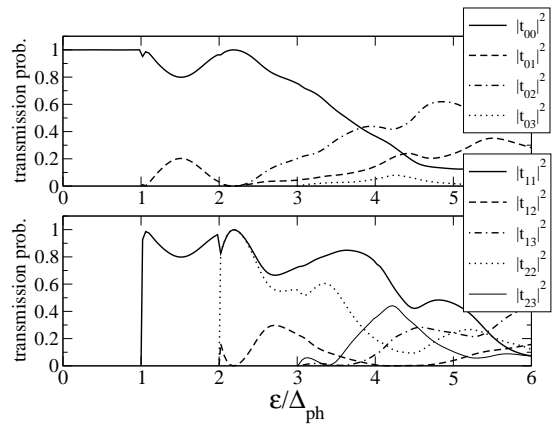}
\caption{\label{Tij-ph-R2.0 figure}Individual phonon transmission probabilities $|t_{nn'}|^2$ as a function of energy for a circular right-angle bend with $R=2 \, b$. Here $\Delta_{\rm ph} \equiv \pi \hbar v / b$.}
\vskip 0.1in
\end{figure}

\begin{figure}
\includegraphics[width=7.2cm]{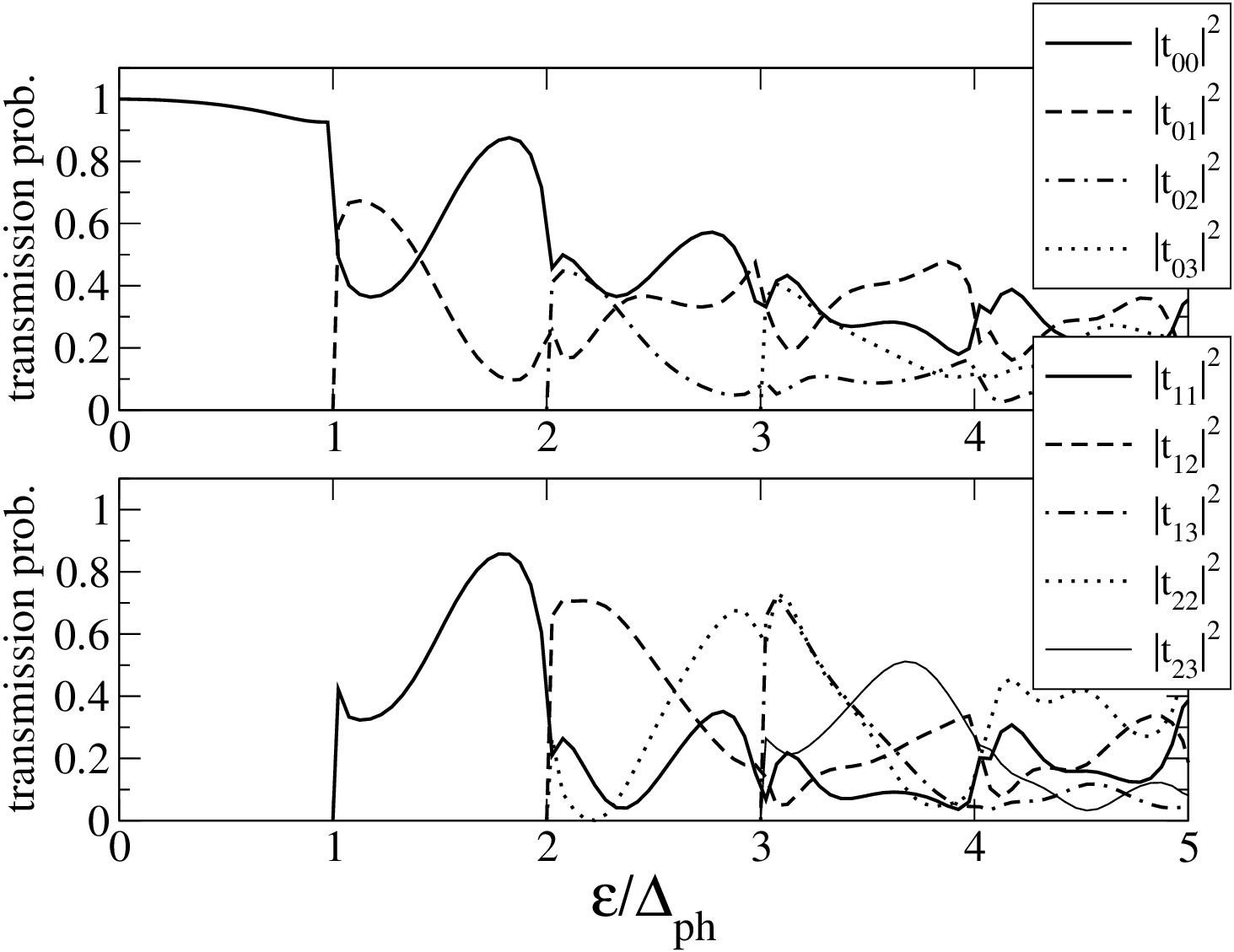}
\caption{\label{Tij-ph-R1.001 figure}Individual phonon transmission probabilities $|t_{nn'}|^2$ as a function of energy for a circular right-angle bend with $R=1.001 \, b $.}
\end{figure}

In Fig.~\ref{Tij-ph-R2.0 figure} we give the individual phonon transmission probabilities $|t_{nn'}(\epsilon)|^2$, defined in Eq.~(\ref{transmission matrix}), for the lowest channels in a smoothly bent wire with $R=2 \, b$. In Fig.~\ref{Tij-ph-R1.001 figure} we do the same for a more tightly bent wire, with $R=1.001 \, b $ (inner radius is $10^{-3} \, b$). At energies given by
\begin{equation}
\epsilon = n \Delta_{\rm el}  \ \ \ \ \ {\rm with} \ \ \ \ \  n=1,2,3,\dots,
\label{phonon threshold energies}
\end{equation}
where $\Delta_{\rm ph} \equiv \pi \hbar v / b$, additional channels in the wire become propagating and contribute to thermal transport.

In both examples, transmission is nearly perfect in the low-energy $\epsilon <  \Delta_{\rm ph}$ limit, where there is only a single propagating channel. At higher energies, scattering does occur. However it is mostly in the forward direction, and the fraction of transmitted energy ${\sf T}_{\! \rm ph}(\epsilon)$, defined in Eq.~(\ref{phonon probability definition}), is essentially unchanged from that of a straight wire.

\begin{figure}
\includegraphics[width=7.2cm]{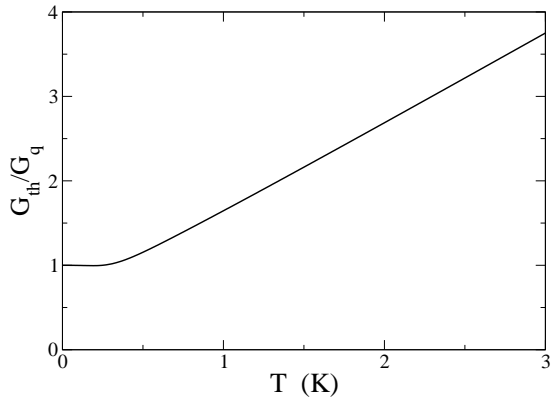}
\caption{\label{Gth-R1.001-b100nm figure} Dimensionless thermal conductance $G_{\rm th}/G_{\rm q}$ as a function of temperature, for a $100 \, {\rm nm}$ curved Si-like quantum wire, with outer radius $R =1.001 \, b$. Here $G_{\rm q} \equiv {\pi k_{\rm B}^2 T/6 \hbar}$, which is itself linearly proportional to temperature.}
\vskip 0.2in
\end{figure}

\begin{figure}
\includegraphics[width=7.2cm]{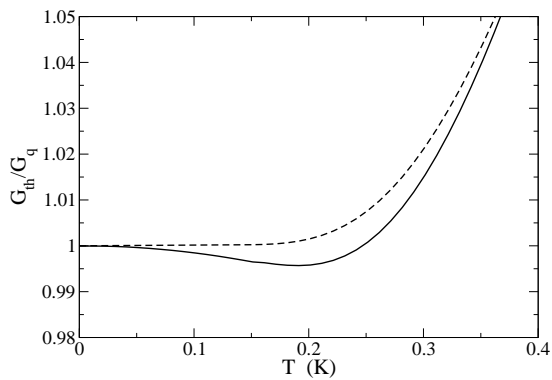}
\caption{\label{Gth vs Gq figure} The solid curve is the same as Fig.~\ref{Gth-R1.001-b100nm figure}. Dashed curve is the dimensionless thermal conductance for a straight Si-like wire with $b =100 \, {\rm nm}$. Thermal transport is hardly suppressed by the bending.}
\end{figure}

In Fig.~\ref{Gth-R1.001-b100nm figure} we plot the thermal conductance $G_{\rm th}$ in units of the ``quantum'' of thermal conductance 
\begin{equation}
G_{\rm q} \equiv {\pi k_{\rm B}^2 T \over 6 \hbar} \approx 0.95 \, T \ {\rm pW \, K^{-2}},
\label{thermal conductance quantum}
\end{equation}
for a curved wire of width $b =100 \, {\rm nm}$ and outer radius $R = 1.001 \, b$. The scalar phonon velocity is taken to be $v = 8.5 \! \times \! 10^{5} \, {\rm cm \ s^{-1}}$, the longitudinal sound speed in Si. The thermal transport is hardly affected by the curvature in the wire, as can be seen in Fig.~\ref{Gth vs Gq figure}, which compares an expanded plot of $G_{\rm th}/G_{\rm q}$ to that for a straight wire. The greatest suppression occurs near  $200 \, {\rm mK}$ and is only about $0.5 \%$ of the thermal conductance quantum.

\begin{figure}
\includegraphics[width=7.2cm]{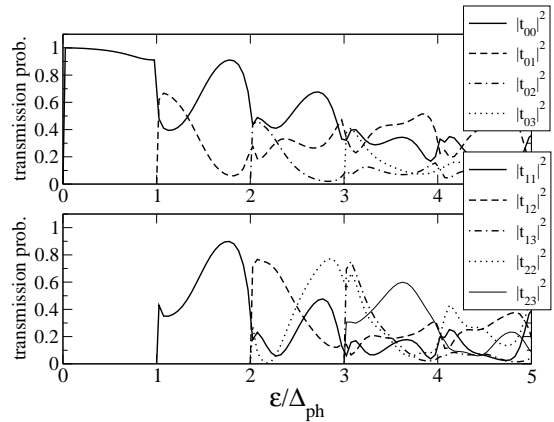}
\caption{\label{Tij-ph-R1.00001 figure}Individual phonon transmission probabilities $|t_{nn'}|^2$ as a function of energy for a circular right-angle bend with $R=1.00001 \, b$.}
\end{figure}

It is physically unrealistic to consider a $100 \, {\rm nm}$ wire bent more sharply than $R = 1.001 \, b$, because the inner radius of $0.10 \, {\rm nm}$ in this case is already approaching atomic dimensions. However, for a wire of width $b = 10 \, \mu{\rm m}$ and the same inner radius of curvature, we have $R = 1.00001 \, b$, the transmission probabilities for which are shown in Fig.~\ref{Tij-ph-R1.00001 figure}. The transmission probabilities when $R = 1.00001 \, b$ are similar to that for $R = 1.001 \, b$, shown previously in Fig.~\ref{Tij-ph-R1.001 figure}, as is the thermal conductance. We find in that in a $10 \, \mu{\rm m}$ Si-like quantum wire with $R = 1.00001 \, b$, the greatest suppression in $G_{\rm th}$ occurs near $2 \, {\rm mK}$ and is again only about $0.5 \%$ of the conductance quantum in magnitude.

\section{DISCUSSION}
\label{discussion section}

We have introduced a general method to calculate the transmission of scalar waves appropriate for mesoscopic electron and phonon transport through a curved wire or waveguide. Applications to electron transport accurately reproduce results obtained by other methods. Phonon transport through curved wires is considered here for the first time.

Our results demonstrate that curvature hardly suppresses thermal transport, even for sharply bent wires, at least within the two-dimensional scalar phonon model considered. This behavior can, to some extent, be understood by considering transport in the extreme long- and short-wavelength limits. In the long-wavelength, low energy limit, ${\sf T}_{\! \rm ph} \rightarrow 1$, a consequence of the rigid-body nature of the wire. ${\sf T}_{\! \rm ph}$ also approaches unity for short wavelengths, because in this limit the phonons cannot sense the curvature.

Because the phonon reflection probability always vanishes in the long-wavelength limit, a simple perturbative (Born approximation) treatment is posssible at low energies. For example, the energy-dependent $n=0$ transmission probability is
\begin{equation}
|t_{00}|^2 = 1-|r_{00}|^2,
\label{Born transmission coefficient}
\end{equation}
where, to leading nontrivial order,
\begin{eqnarray}
r_{00} &=& - {i \hbar v \over 2 \epsilon b}  \int \! d^2R \, d^2R' \ e^{i \epsilon X/ \hbar v} \nonumber \\
&\times& V_{\! \scriptscriptstyle R} \, G_0({\bf R},{\bf R}', \epsilon) \, V_{\! \scriptscriptstyle {R^\prime}} \, e^{i \epsilon X'/\hbar v}.
\label{Born reflection coefficient}
\end{eqnarray}
$V_{\! \scriptscriptstyle R}$ and $V_{\! \scriptscriptstyle {R^\prime}}$ act on ${\bf R}$ and ${\bf R}^\prime$, respectively. This result allows the low-temperature thermal transport though a variety of wire shapes to be addressed quite simply. Although an analogous perturbative expression can be derived for the electronic transmission probability as well, the form of the transverse part $\chi_n$ of the unperturbed scattering states, as dictated by the hard-wall boundary conditions, then leads to a divergence in the Born series,\cite{electron born limit} consistent with the fact that ${\sf T}_{\rm el} \rightarrow 0$ in the long-wavelength limit.

We conclude with a brief discussion of the experimental implications of our mesoscopic thermal transport results, the charge-trasport case having already been discussed in the literature.\cite{Londergan book,Baranger} Thermal transport in carbon nanotubes has been studied experimentally by several groups,\cite{Hone etal nanotubes,Kim etal nanotubes,Yang etal nanotubes} and nanotubes would be interesting systems to use to investigate the effects of bending on transport. To apply our method of analysis to this system would require the consideration of scattering of elastic waves in a curved, hollow tube. Although they were obtained for scalar phonons in two-dimensional strips, our results do suggest that the effects of curvature will be small, if not completely negligible, in these systems.

\acknowledgments

This work was supported by the National Science Foundation under CAREER Grant No.~DMR-0093217, and by the Research Corporation. Acknowledgment is also made to the Donors of the American Chemical Society Petroleum Research Fund, for partial support of this research. It is a pleasure to thank Miles Blencowe, Andrew Cleland, Dennis Clougherty, and Kelly Patton for useful discussions, and Aleksandar Milo\v{s}evi\'c for his contributions at the beginning stages of this work.

\appendix

\section{INTEGRALS INVOLVING PRODUCTS OF DIRAC DELTA FUNCTIONS AND STEP FUNCTIONS}
\label{delta function appendix}

Here we discuss integrals of the form
\begin{equation}
\int_{-\infty}^{\infty} \!  dx \, F[\Theta(x)] \, \delta(x),
\label{undefined integral}
\end{equation}
where $F$ is twice continuously differentiable, and $\Theta(x)$ and $\delta(x)$ are the unit step and Dirac delta functions, respectively. Integrals of the form (\ref{undefined integral}), which involve {\it products} of generalized functions, depend sensitively on how $\Theta(x)$ and $\delta(x)$ are defined. It will be necessary to define $\Theta(x)$ and $\delta(x)$ according to their appearance in this work.

The step function $\Theta(x)$ appearing in Eq.~(\ref{undefined integral}) originated from the curvature profile of Eq.~(\ref{right angle profile}) used to describe a circular segment of wire connected to a straight lead, and the delta function comes from its derivative with respect to arclength  in Eq.~(\ref{right angle profile derivative}), which is required by the effective potential $V$ of Eq.~(\ref{effective potential}). Therefore we require $\Theta(x)$ to be smooth (on some microscopic scale) and continuous, and $\delta(x)$ to be related to it by
\begin{equation}
{d \over dx} \Theta(x) = \delta(x).
\end{equation}
We also require, of course, that 
\begin{equation}
\lim_{x \rightarrow -\infty} \Theta(x) = 0 \ \ \ \ \ {\rm and} \ \ \ \ \  \lim_{x \rightarrow \infty} \Theta(x) = 1.
\end{equation}
The precise shape of $\Theta(x)$ near $x=0$ is immaterial, but with no loss of generality we can require that $\Theta(0) = {\textstyle{1 \over 2}}$.

Integrals of the form (\ref{undefined integral}) are now well defined. For example,
\begin{equation}
\int_{-\infty}^{\infty} \!  dx \, \Theta(x) \, \delta(x) = {\textstyle{1 \over 2}},
\label{first example integral}
\end{equation}
as expected, but
\begin{equation}
\int_{-\infty}^{\infty} \!  dx \, [\Theta(x)]^2 \, \delta(x) = {\textstyle{1 \over 3}}, 
\label{second example integral}
\end{equation}
instead of ${\textstyle{1 \over 4}}$. These results are obtained by integrating by parts and using the behavior of $\Theta(x)$ as $x \rightarrow \pm \infty$, {\it not} by evaluating $\Theta(0)$ and $[\Theta(0)]^2.$ More generally,
\begin{equation}
\int_{-\infty}^{\infty} \!  dx \, [\Theta(x)]^n \, \delta(x) = {1 \over n+1},  \ \ \ \ \ ({\rm for} \ n>0)
\label{third example integral}
\end{equation}
which is different from the naive value of $[\Theta(0)]^n = ({\textstyle{1 \over 2}})^n,$ unless $n=1$.

The reason why 
\begin{equation}
\int_{-\infty}^{\infty} \!  dx \, F[\Theta(x)] \, \delta(x) \neq F[\Theta(0)]
\end{equation}
in some of these examples is because the delta function is distributed over a small but finite region of $x$, whereas $\Theta(x)$ and $F[\Theta(x)]$ are generally {\it not} slowly varying over that length scale. We conclude, therefore, that integrals of the form (\ref{undefined integral}) appearing in the evaluation of matrix elements of $V$, have to be evaluated using integration-by-parts (or with an equivalent method).


\begin{thebibliography}{99}
\bibitem{thickness footnote} The two-dimensional Helmholtz equation is valid at energy scales considerably less than $v / d$, where $d$ is the thickness of the wire in the $z$ direction.
\bibitem{constant width footnote} Our method, of course, has it's own limitations. For example, in the present formulation the width of the wire is assumed to be constant.
\end{thebibliography}
\end{document}